\documentclass{article}

\usepackage{arxiv}

\usepackage[utf8]{inputenc} 
\usepackage[T1]{fontenc}    
\usepackage{hyperref}       
\usepackage{url}            
\usepackage{booktabs}       
\usepackage{amsfonts}       
\usepackage{nicefrac}       
\usepackage{microtype}      
\usepackage{lipsum}		
\usepackage{graphicx}
\usepackage{natbib}
\usepackage{doi}

\usepackage{blindtext}
\usepackage{algorithm}
\usepackage{algpseudocode}
\usepackage{enumerate}

\algrenewcommand\algorithmicrequire{\textbf{Input:}}
\algrenewcommand\algorithmicensure{\textbf{Output:}}
\usepackage{lastpage}

\algdef{SE}[SUBALG]{Indent}{EndIndent}{}{\algorithmicend\ }%
\algtext*{Indent}
\algtext*{EndIndent}

\title{Parallel Graph Drawing Algorithm for Bipartite Planar Graphs}

\author{{\includegraphics[scale=0.06]{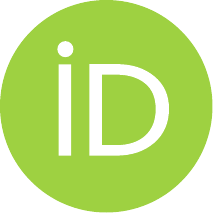}\hspace{1mm}Naman Jain} \\
	Student\\
	The International School Bangalore\\
	Bangalore, India \\
	\texttt{jnaman@tisb.ac.in} \\
}
\date{}
\renewcommand{\headeright}{}
\renewcommand{\undertitle}{}
\renewcommand{\shorttitle}{}
\hypersetup{
pdftitle={Parallel Graph Drawing Algorithm for Bipartite Planar Graphs},
pdfsubject={},
pdfauthor={Naman Jain},
pdfkeywords={Computational Complexity, Parallel Algorithms, Graph Drawing},
}

\begin{document}
\maketitle

\begin{abstract}
We give a parallel $O(\log(n))$-time algorithm on a CRCW PRAM to assign vertical and horizontal segments to the vertices of any planar bipartite graph $G$ in the following manner:
i) Two segments cannot share an interior point
ii) Two segments intersect if and only if the corresponding vertices are adjacent, which uses a polynomial number of processors. In other words, represent vertices of a planar bipartite graph as parallel segments, and edges as intersection points between these segments. Note that two segments are not allowed to cross. Our method is based on a parallel algorithm for st-numbering which uses an ear decomposition search.
\end{abstract}


\section{Introduction}
Computational Complexity Theory is the study of the amount of resources required by algorithms to solve computational problems. In particular, it focuses on the time and space used by an algorithm. It is an important field of theoretical computer science as it offers a framework to understand and compare the efficiency of algorithms, and find the most suitable solutions. Furthermore, it allows us to classify problems using complexity classes such as $\mathsf{P}$, $\mathsf{NP}$, and $\mathsf{NC}$.\\\newline
While working on complex computational problems, the primary focus is to optimize algorithms for efficiency and speed. \cite{DBLP:journals/dcg/FraysseixMP95} introduced an algorithm to assign segments to the vertices of planar graphs in the research paper ``A Left-First Search Algorithm for Planar Graphs". While this Left-First Search (LFS) algorithm offered a satisfactory time period for sequential algorithms, in today’s world of big data and faster processing, we aim for even faster parallel solutions using multiple processors.

\subsection{Motivation}
Graph drawing is a field of mathematics in which we explore visualizations of the information that a graph contains. It aims to offer a representation of a graph in which all its information and structure is easy to understand (\cite{DBLP:books/ws/NishizekiR04}). Moreover, it may be used to create visually appealing graphs.
The addressed problem belongs to the field of graph drawing. It has connections to visibility, incidence relations between geometry objects and more.
\\\\
We aim to parallelize the algorithm proposed by \cite{DBLP:journals/dcg/FraysseixMP95}. Parallel algorithms are currently at the forefront of computational complexity theory. We now have parallel computers which contain massive numbers of processors so, we require parallel algorithms to efficiently use these. Parallel algorithms aim to be significantly faster than sequential algorithms, with a polylogarithmic time parallel algorithm considered acceptable. As a result, parallel algorithms are often more relevant practically. We aim to transform this algorithm into a parallel $O(\log{n})$ time algorithm on a CRCW PRAM using a polynomial number of processors.

\subsection{Problem Statement}

A planar graph is defined as one which can be embedded onto a plane without its edges crossing. A bipartite graph is a graph for which the vertices can be divided into two disjoint sets, with no adjacent vertices within the same set. In this paper, we exclusively consider finite graphs $G$ without loops, but multiple edges are allowed. We are looking for a parallel algorithm to assign vertical and horizontal segments to the vertices of any planar bipartite graph $G$ in the following manner:
\begin{enumerate}
    \item Two segments cannot share an interior point, or cross.
    \item Two segments intersect if and only if the corresponding vertices are adjacent.
\end{enumerate}

\subsection{Statement of Result}

We describe a parallel $O(\log{n})$ time algorithm on a CRCW PRAM. It uses $n^{O(1)}$ processors to assign segments to the vertices of a graph, in the manner previously described, in polylogarithmic time.\\
\newline We begin by quadrilateralization of the bipartite planar graph. Next, we use the idea of st-numbering to convert the graph into a directed, acyclic graph. We use a parallel algorithm for st-numbering created by \cite{DBLP:journals/tcs/MaonSV86}. This algorithm uses an ``ear decomposition search". Lastly, we use this st-numbering to assign the segments to the vertices of a bipartite planar graph $G$. The concept of st-numbering is defined in the next section, alongside an explanation of the ear decomposition search and parallel algorithm to find an st-numbering. All the subroutines within this algorithm are within parallel $O(\log{n})$ time algorithm on a CRCW PRAM so, the entire algorithm is a parallel $O(\log{n})$ time algorithm on a CRCW PRAM.

\subsection{Related Work}
A parallel planarity testing algorithm, in $O(\log{n})$ time, is presented by \cite{DBLP:journals/jcss/RamachandranR94}. This algorithm creates an embedding of the input graph on a plane if this is possible, or else outputs that the graph is non-planar. This algorithm too makes use of the technique of open ear decomposition.
\subsection{Organization of the Paper}
The rest of the paper is organized as follows. Section 2 introduces preliminaries, which are relevant previous results and definitions. Section 3 describes the existing sequential algorithm for the same problem. Section 4 describes a parallel algorithm for quadrilateralization of a planar bipartite graph. Section 5 contains the entire proposed parallel algorithm. Here, we also prove its correctness and resource bounds. Section 6 is the conclusion, with a short summary of the results.

\section{Preliminaries}
Given a graph $G=(V,E)$, and 2 vertices $s,t\in V$, an $st$-numbering of $G$ is a numbering of the vertices of graph $G$ such that $s$ gets the lowest number, $t$ gets the highest number and all the other vertices are adjacent to another vertex with a lower and higher number.

Alternatively, given an edge $\vec{e}= \vec{st}$, an oriented graph $\vec{G}$ has a valid st-numbering if:
\begin{enumerate}
    \item $s$ and $t$ are the only source and sink, respectively, in $\vec{G}$
    \item $\vec{G}$ is acyclic
\end{enumerate}
An open ear decomposition requires a graph $G=(V,E)$ and an edge $e=(s,t)$ which is part of the edge set of $G$. An ear decomposition of $G$, starting with $P_0$ is defined as a decomposition $E=P_0\cup P_1\cup \cdots \cup P_k$, where $P_{i+1}$ is a simple path whose endpoints lie on $P_0\cup P_1\cup \cdots \cup P_i$, but interval vertices do not(\cite{DBLP:journals/tcs/MaonSV86}).\\
An ear decomposition is categorised as an open ear decomposition if the endpoints of each path $P_i$ do not coincide.\\
\newline In a planar bipartite graph, $G=((U,V),E)$ let $u_0,u_1,\cdots,u_n$ be the ``blue vertices", or vertices in $U$ with no edges between them. Let $v_0,v_1,\cdots,v_n$ be the ``red vertices", or vertices in $V$ with no edges between them. A bipartite planar graph is defined as a quadrilateralization if each of its faces contains 4 edges and there are no multiple edges (\cite{DBLP:journals/dcg/FraysseixMP95}).\\
\newline In the parallel algorithm presented in this paper, we make use of a parallel algorithm to produce a valid st-numbering presented by \cite{DBLP:journals/tcs/MaonSV86}. This algorithm takes in the following input:
\begin{enumerate}
    \item A graph $G(V,E)$ and a specific edge $(s,t)\in E$
    \item An open ear decomposition
\end{enumerate}
to produce a valid st-numbering of $G$.\\
\newline In total, the algorithm takes $O(\log{n})$ time using $ ((n \log{n} + m)/\log{n})$ processors, where $n$ and $m$ are the number of vertices and edges in the graph, respectively. This algorithm runs on a CRCW (Concurrent-Read Concurrent-Write) PRAM (Parallel Random Access Machine).\\
\newline This algorithm has two major sections. First, the edges of the graph are oriented to form a directed graph which is acyclic, and where each vertex lies on some path from $s$ to $t$ (Stage 1). Next, a topological sort is conducted upon this directed graph to assign the numbers 1 to $n$ to its vertices (Stage 2). \\
\newline This is conducted using the ear decomposition search, presented by \cite{DBLP:journals/tcs/MaonSV86}.\\
\newline Another important preliminary in this paper is the justification of the sequential algorithm presented by \cite{DBLP:journals/dcg/FraysseixMP95} for the assignment of segments, which is explored below. 

\section{Brief Summary of Existing Sequential Algorithm}
In the paper by \cite{DBLP:journals/dcg/FraysseixMP95}, a sequential algorithm is presented for the same segment assignment. Here, using a ``na\"ive linear time algorithm", the bipartite planar graph is extended to a quadrilateralization. Next, a sequential greedy algorithm is used to assign a valid st-numbering to the vertices of a bipartite planar graph, $G$. For a quadrilateralization $G$, whose vertices are colored in blue and red, connect the 2 red vertices and the 2 blue vertices in a face $f$ from an edge passing through $f$. The graph $G_u$ formed by these edges is the graph of blue diagonals and $G_v$ is the graph of red diagonals. Let $s_v,s_u,t_v,t_u$ denote the vertices of the outer face of $G$.\\
\newline As described by \cite{DBLP:journals/dcg/FraysseixMP95}, we use this to find an $s_ut_u$-numbering of the blue vertices. Once this is done, we have the red vertices ordered as $(v_1,v_2,\cdots v_p)$ and the blue vertices ordered as $(u_1,u_2,\cdots u_q)$. Now, we assign the segments. For all the red vertices $v_i\, |\,1\leq i \leq p$, we assign a vertical segment $V_i$, with the endpoints $(i, min_{\{{v_i},{u_j}\}\in G}\,j)$ and $(i, max_{\{{v_i},{u_j}\}\in G}\,j)$. Similarly, for all blue points $u_j\,|\,1\leq j \leq q$, we assign a horizontal segment $H_j$, with the endpoints $(min_{\{{v_i},{u_j}\}\in G}\,i,j)$ and $(max_{\{{v_i},{u_j}\}\in G}\,i,j)$. As proven by \cite{DBLP:journals/dcg/FraysseixMP95}, these segments meet the requirements defined above.
\section{Parallel Quadrilateralization Algorithm}
We start off by converting our bipartite planar graph into a quadrilateralization using a parallel algorithm. For this algorithm, we input a sorted incident list of vertices in each face. The list is sorted by Cole's Sorting Algorithm (\cite{DBLP:journals/siamcomp/Cole88}).
\begin{algorithm}[H]
\caption{A Parallel Algorithm to Quadrilateralize a Bipartite Planar Graph}\label{alg:cap}
\begin{algorithmic}
\Require Incident lists of red ($(v_1,v_2\cdots v_q)$) and blue ($(v_1,v_2\cdots v_q)$) vertices of each face $f$ in a Bipartite Planar Graph $G$.
\Ensure Quadrilateralization $\tilde{G}$\\
\State $\mathbf{parfor}$ each face $f$ $\mathbf{do}$:
\Indent
\State Take the first red vertex ($v_1$) from the sorted incidence list of $f$
\State $\mathbf{parfor}$ each blue vertex $u_i$ $\mathbf{do}$:
\Indent
\State Output the edge $(u_i,v_1)$
\EndIndent
\EndIndent
\end{algorithmic}
\end{algorithm}
In the above algorithm, one processor is assigned to each blue vertex, which then outputs an edge between that vertex, and the red vertex with the lowest number in that face, in $O(1)$ time. Hence, each edge will be added simultaneously by a different processor. We now investigate the time complexity of this algorithm.\\\newline 
First, we use Cole's sorting algorithm, which takes $O(\log{n})$ time. We then select the first red vertex from the sorted incident list of each face. Let this vertex be $v_0$. One processor is assigned to each blue vertex, with the aim of outputting an edge between that vertex and $v_0$. Let $u_1$ and $u_n$ be the 2 blue vertices adjacent to $v_0$. From the clockwise ordering of vertices in face $F$, delete vertex $v_0$ and then extract the clockwise order of vertices between $u_1$ and $u_n$. Now, the aim is to splice the list of vertices between $u_1$ and $u_n$ (face rotation scheme) into the list containing vertices adjacent to $v_0$ (vertex rotation scheme). For this, the pointer of each $u_k$ vertex points to the vertex which is a 2-length clockwise path away. Similarly, we create a chain of points $u_k$. Now, the pointer at $u_1$ is changed from $u_n$ to $u_2$, and the pointer coming to $u_n$ is changed from $u_1$ to $u_{n-1}$, on a doubly-linked link list, as seen in the picture below.
\begin{figure}[h!]
    \centering
    \includegraphics[width=15cm]{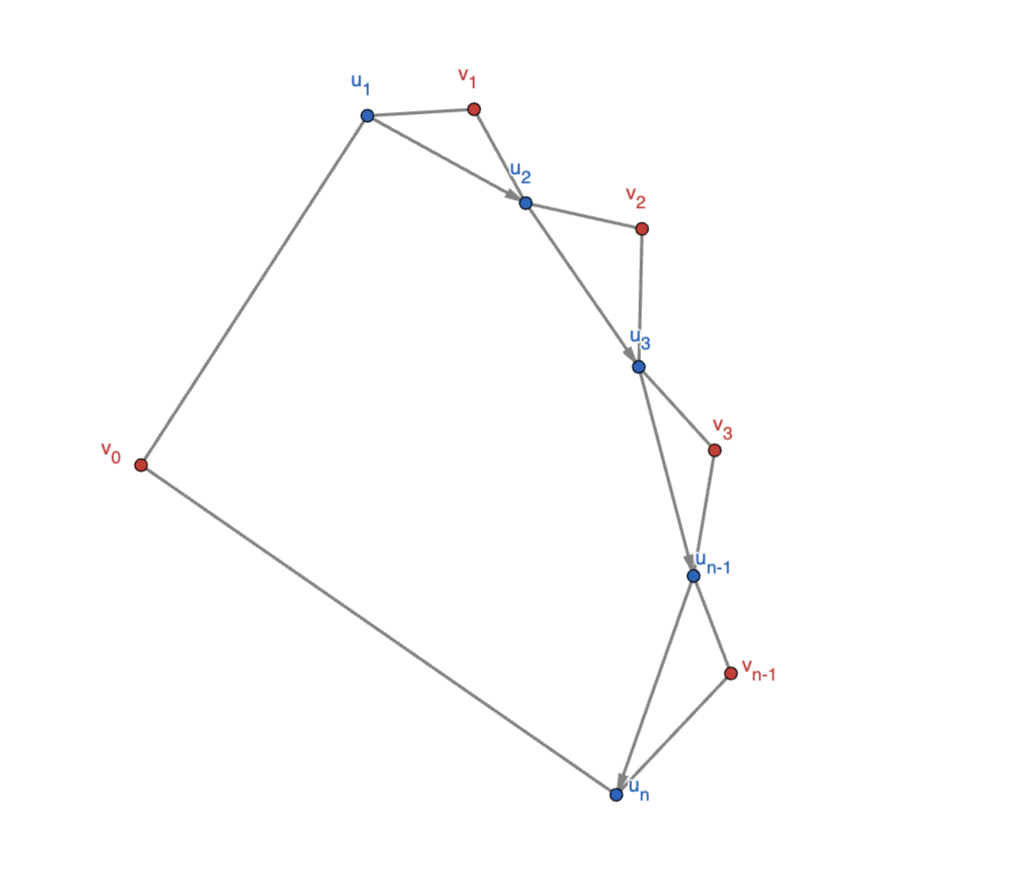}
    \caption{An image showing the corrected pointers from each point $u_k$}
\end{figure}

As a result, we can splice the lists together and add all the necessary edges to $v_0$'s adjacency list in $O(1)$ time.\\
\clearpage
\begin{figure}[h!]
    \centering
    \includegraphics[width=15cm]{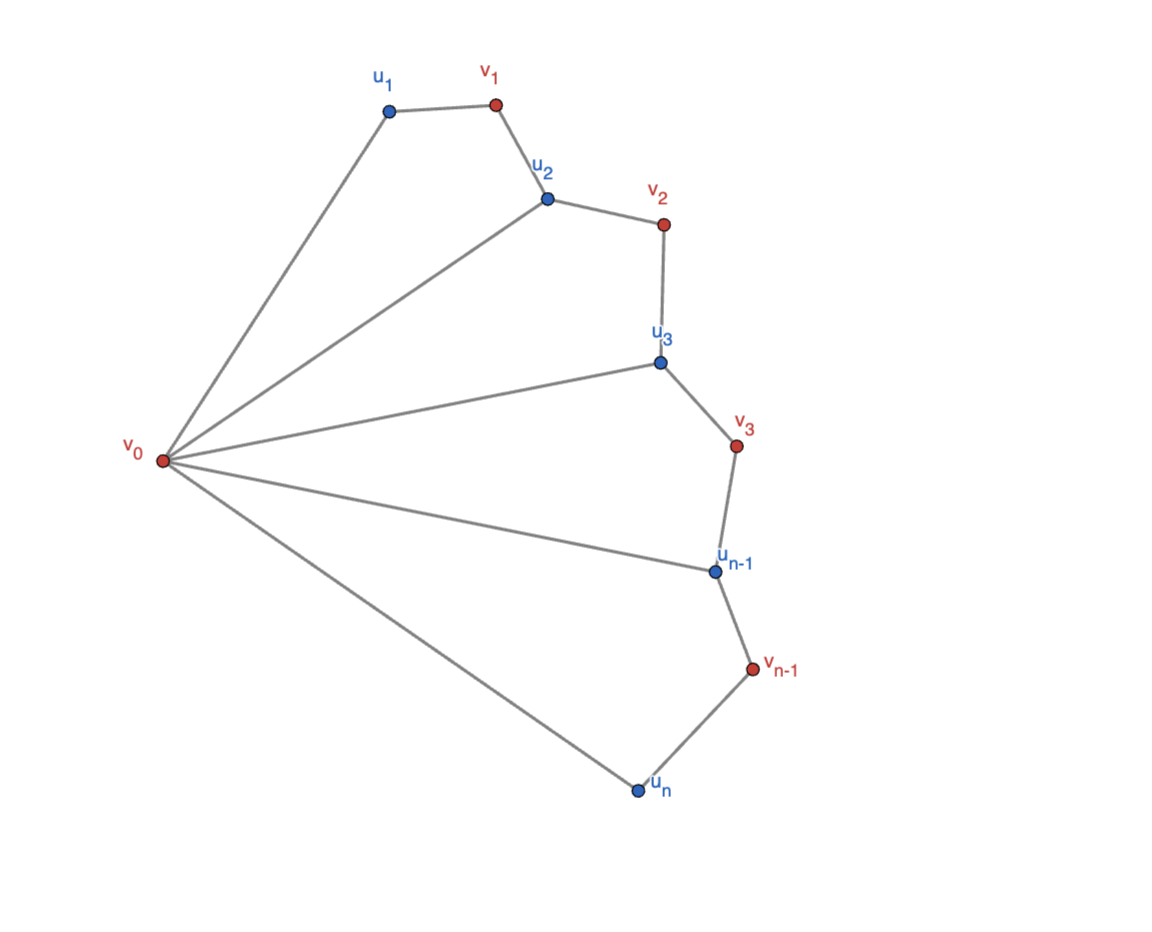}
    \caption{An image showing the graph with new edges added}
\end{figure}
Since it takes $O(\log{n})$ time to find the minimum vertex and carry out other parts in the algorithm, the time period of the parallel algorithm to quadrilateralize a bipartite planar graph is $O(\log{n})$. The number of processors required is equivalent to the number of face-vertex pairs, which is equivalent to the number of angles in a graph. The number of angles is equal to twice the number of edges at each vertex. Hence, the number of processors is linear in $m$, and hence is also linear in $n$.
\section{Entire Parallel Algorithm to Assign Segments}
We then find a valid st-numbering for the vertices of the graph using the parallel algorithm by \cite{DBLP:journals/tcs/MaonSV86} described in Preliminaries.\\
\newline After obtaining a valid st-numbering and quadrilateralization of the bipartite planar graph, the rest of the algorithm follows the same steps as the sequential algorithm. \\
\newline We assign the segments in the way described by \cite{DBLP:journals/dcg/FraysseixMP95}. For all the red vertices $v_i\, |\,1\leq i \leq p$, we assign a vertical segment $V_i$, with the endpoints $(i, min_{\{{v_i},{u_j}\}\in G}\,j)$ and $(i, max_{\{{v_i},{u_j}\}\in G}\,j)$. Similarly, for all blue points $u_j\,|\,1\leq j \leq q$, we assign a horizontal segment $H_j$, with the endpoints $(min_{\{{v_i},{u_j}\}\in G}\,i,j)$ and $(max_{\{{v_i},{u_j}\}\in G}\,i,j)$. This segment assignment meets all the requirements from the problem statement, as proven by \cite{DBLP:journals/dcg/FraysseixMP95}.
\begin{algorithm}[H]
\caption{Entire Parallel Algorithm to Assign Segments to the Vertices}\label{alg:cap}
\begin{algorithmic}
\Require Incident lists of red ($(v_1,v_2\cdots v_q)$) and blue ($(v_1,v_2\cdots v_q)$) vertices of each face $f$ in a Bipartite Planar Graph $G$.
\Ensure Segment Assignment, as described in the problem statement.\\
\State 1. Use Algorithm 1 to quadrilateralize $G$
\State 2. Find a valid st-numbering for quadrilateralisation $\tilde{G}$ using the ear-decomposition search (\cite{DBLP:journals/tcs/MaonSV86}) 
\State 3. Use st-numbering to assign segments to the vertices of $\tilde{G}$ as described above.
\end{algorithmic}
\end{algorithm}
Step 1 requires $O(1)$ time using $m$ processors. The other subroutines within this algorithm run in parallel $O(\log{n})$ time on a CRCW PRAM so, the entire algorithm is a parallel $O(\log{n})$ time algorithm on a CRCW PRAM.

\section{Conclusion}
To conclude, we have found a parallel $O(\log{n})$ time algorithm  on a CRCW PRAM for this graph-drawing problem. 
\section*{Acknowledgements}
I would like to thank Dr Samir Datta, Professor of Computer Science at Chennai Mathematical Institute, India for all his guidance and support in this paper. It would not have been possible without him. I would also like to thank my teachers, parents and sister for their continuous support.
\bibliographystyle{unsrtnat}
\bibliography{references}  






\end{document}